\documentclass{article}%
\usepackage{amsmath}
\usepackage{amssymb}
\usepackage{epsfig}
\usepackage{psfig}
\usepackage{latexsym}
\usepackage{amsfonts}
\usepackage{graphicx}%
\begin{document}
\parindent 0mm \setlength{\parskip}{\baselineskip} \thispagestyle{empty}
\pagenumbering{arabic} \setcounter{page}{0} \mbox{ }
\rightline{UCT-TP-265/05}\newline\rightline{November 2005}\newline%
\vspace{0.2cm}

\begin{center}
{\Large \textbf{Chiral condensates from tau}} {\Large -\textbf{decay: a
critical reappraisal}}{\LARGE \footnote{{\LARGE {\footnotesize Supported by
MCYT-FEDER under contract FPA2002-00612, EC-RTN under contract HPRN-CT2000/130
and partnership Mainz-Valencia Universities, and DFG-NRF agreements.}}}}

\textbf{Jos\'{e} Bordes}$^{(a)},$ \textbf{Cesareo A. Dominguez$^{(b)}$,
Jos\'{e} Pe\~{n}arrocha}$^{(a)}$, \textbf{Karl Schilcher$^{(c)}$}$\bigskip$

$^{(a)}$Departamento de F\'{\i}sica Te\'{o}rica-IFIC, Universitat de
Valencia-CSICE-46100 Burjassot-Valencia, Spain

$^{(b)}$Institute of Theoretical Physics and Astrophysics\\[0pt]University of
Cape Town, Rondebosch 7700, South Africa

$^{(c)}$Institut f\"{u}r Physik, Johannes Gutenberg-Universit\"{a}t\\[0pt]%
Staudingerweg 7, D-55099 Mainz, Germany\newline\vspace{0.2cm}
\textbf{Abstract}
\end{center}

\noindent The saturation of QCD chiral sum rules is reanalyzed in view of the
new and complete analysis of the ALEPH experimental data on the difference
between vector and axial-vector correlators (V-A). Ordinary finite energy sum
rules (FESR) exhibit poor saturation up to energies below the tau-lepton mass.
A remarkable improvement is achieved by introducing pinched, as well as
minimizing polynomial integral kernels. Both methods are used to determine the
dimension $d=6$ and $d=8$ vacuum condensates in the Operator Product
Expansion, with the results: $\mathcal{O}_{6}(2.6\; \mbox{GeV}^{2})=-(0.00226\pm
0.00055)\;\mbox{GeV}^{6}$ , and $\mathcal{O}_{8}(2.6\; \mbox{GeV}^{2})=-(0.0054\pm
0.0033)\;\mbox{GeV}^{8}$ from pinched FESR, and compatible values from the
minimizing polynomial FESR. Some higher dimensional condensates are also
determined, although we argue against extending the analysis beyond dimension
$d=8$. The value of the finite remainder of the (V-A) correlator at zero
momentum is also redetermined: $\bar{\Pi}(0)=-4\;\bar{L}_{10}=0.02579\pm
0.00023$. The stability and precision of the predictions are significantly
improved compared to earlier calculations using the old ALEPH data. Finally,
the role and limits of applicability of the Operator Product Expansion in this
channel are clarified.

\bigskip\noindent

\section{Introduction}

\noindent More than twenty five years ago, Shifman, Vainshtein and Zakharov
\cite{SVZ} proposed to use the Operator Product Expansion (OPE) in hadronic
current-current correlators to extend asymptotic predictions of QCD to low
energies. In this approach there appear universal vacuum expectation values of
quark and gluon fields, the so-called vacuum condensates, which have to be
extracted from experiment. This extraction is usually carried out by using
methods based on dispersion relations. Ultimately, one has to relate error
afflicted data in the time-like region to asymptotic QCD in the space-like
region. Unfortunately, this task of analytic continuation constitutes,
mathematically, a so-called ill-posed problem. In fact, extracting condensates
from data is highly sensitive to data errors. Not surprisingly, results from
different collaborations have not been always consistent \cite{Friot}. The
main reason for these inconsistencies was frequently the impossibility of
estimating reliably the errors in the method. With the release of the final
analysis of the precise measurements of the vector (V) and axial-vector (A)
spectral functions obtained from tau-lepton decay by the ALEPH collaboration
\cite{ALEPH2}, an opportunity has been opened to check the validity of QCD sum
rules and the extraction of condensates in the light-quark sector with
unprecedented precision. It is therefore appropriate to reanalyze the data,
taking into account errors and correlations with the least possible
theoretical bias. In this paper we attempt such a critical and conservative
appraisal for the interesting case of chiral sum rules. These sum rules
involve the difference between the vector and the axial-vector correlators
(V-A), which vanishes identically to all orders in perturbative QCD in the
chiral limit. In fact, neglecting the light quark masses, the (V-A) two-point
function vanishes like $1/q^{6}$ in the space-like region, where the scale
$\mathcal{O}(300$ MeV) is set by the four-quark condensates. The interest in
these sum rules is twofold. Apart from describing a QCD order parameter, they
determine the leading contributions of the matrix elements of the electroweak
penguin operators
\begin{align}
Q_{7} &  =6(\bar{s}_{L}\gamma_{\mu}d_{L})\sum\limits_{q=u,d,s}e_{q}(\bar
{q}_{R}\gamma_{\mu}q_{R})\nonumber\\
Q_{8} &  =-12\sum\limits_{q=u,d,s}e_{q}(\bar{s}_{L}q_{R})(\bar{q}_{R}%
d_{L})\nonumber
\end{align}
where $e_{q}$ is the charge of the quark $q$.

In the time-like region, the chiral spectral function $\rho_{V-A}(q^{2})$
should vanish for sufficiently large $Q^{2}\equiv-q^{2}$, but judging from the
ALEPH data \cite{ALEPH2} shown in Fig.1 the asymptotic regime of local duality
does not seem to have been reached, i.e. the spectral function does not vanish
even for the highest momenta attainable in $\tau$-decay.

\begin{figure}
[h]
\begin{center}
\includegraphics[
height=3.0552in,
width=3.0552in
]%
{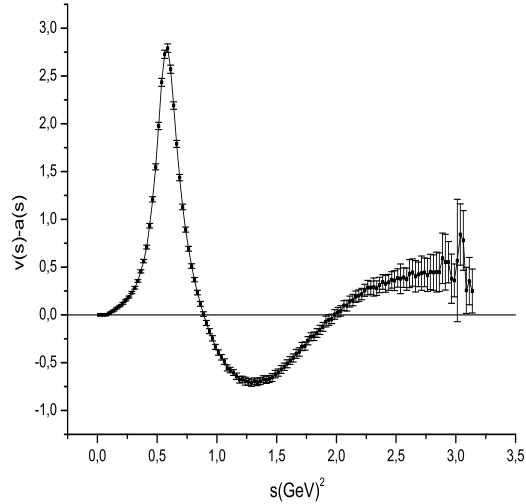}%
\caption{The ALEPH data \cite{ALEPH2} on the vector minus axial-vector
spectral function vs. perturbative QCD (solid line).}%
\label{VmAspec}%
\end{center}
\end{figure}

Under less stringent assumptions, one would hope that at least global duality
should hold in the time-like region. In particular, this should be the case
for the Weinberg-type sum rules \cite{WSR}-\cite{DMO} which involve the first
and second moment of the spectral function. Surprisingly, these sum rules also
appear to be poorly convergent. A likely source of duality violation could be
some non-perturbative contribution to the correlator (e.g. due to instantons)
which falls off exponentially in the space-like region but oscillates in the
time-like region. From Fig.1 it is obvious that convergence could be improved
by incorporating a weight factor which would reduce the non-asymptotic
contribution to the spectral integral. This can be achieved e.g. by
considering so called "pinched sum rules" \cite{Nasrallah} or "minimizing
polynomial sum rules" \cite{BPS}. In view of its importance we have chosen to
reanalyze the issue of duality in chiral sum rules on the basis of the new
ALEPH measurements. Our analysis leads to results showing a significantly
improved accuracy.

\section{Finite energy sum rules}

We begin by defining the vector and axial-vector current correlators
\begin{align}
\Pi_{\mu\nu}^{VV}(q^{2})  &  =i\int d^{4}x \; e^{iqx}<0|T(V_{\mu}(x)V_{\nu
}^{\dagger}(0))|0>\label{2.1}\\
&  =(-g_{\mu\nu}\;q^{2}+q_{\mu}q_{\nu})\;\Pi_{V}(q^{2})\;,\nonumber
\end{align}
\begin{align}
\Pi_{\mu\nu}^{AA}(q^{2})  &  =i\int d^{4}x \;e^{iqx}<0|T(A_{\mu}(x)A_{\nu
}^{\dagger}(0))|0>\label{2.2}\\
&  =\;(-g_{\mu\nu}q^{2}+q_{\mu}q_{\nu})\;\Pi_{A}(q^{2})-q_{\mu}q_{\nu} \;
\Pi_{0}(q^{2})\; ,\nonumber
\end{align}

where $V_{\mu}(x)=:\bar{q}(x)\gamma_{\mu}q(x):$, $A_{\mu}(x)=:\bar{q}%
(x)\gamma_{\mu}\gamma_{5}q(x):$, and $q=(u,d)$. Here we shall concentrate on
the chiral correlator $\Pi_{V-A}\equiv\Pi_{V}-\Pi_{A}$. This correlator
vanishes identically in the chiral limit ($m_{q}=0$), to all orders in QCD
perturbation theory. Renormalon ambiguities are thus avoided. To define our
normalization we note that in perturbative QCD
\begin{equation}
\frac{1}{\pi}\operatorname{Im}\Pi_{V}^{QCD}\left(  s\right)  =\frac{1}{\pi
}\operatorname{Im}\Pi_{A}^{QCD}\left(  s\right)  =\frac{1}{8\pi^{2}}\left(
1+\frac{\alpha_{s}}{\pi} +...\right)  \label{2.3}%
\end{equation}
Non-perturbative contributions due to vacuum condensates contribute to this
two-point function starting with dimension $d=6$, and involving the four-quark
condensate. The Operator Product Expansion (OPE) of the chiral correlator can
be written as

\begin{equation}
\Pi(Q^{2})|_{V-A}=\sum_{N\geq3}^{\infty}\frac{1}{Q^{2N}}\;C_{2N}(Q^{2},\mu
^{2})\;<O_{2N}(\mu^{2})>\;, \label{2.4}%
\end{equation}

with $Q^{2}\equiv-q^{2}$. The scale parameter $\mu$ separates the long
distance non-perturbative effects associated with the condensates $<O_{2N}%
(\mu^{2})>$ from the short distance effects which are included in the Wilson
coefficients $C_{2N}(Q^{2},\mu^{2})$. The OPE is valid for complex $q^{2}$ and
moderately large $|q^{2}|$ sufficiently far away from the positive real axis.
Radiative corrections to the dimension $d=6$ contribution are known
\cite{CH}-\cite{Cirigliano}. They depend on the regularization scheme,
implying that the value of the condensate itself is a scheme dependent
quantity. Explicitly,

\begin{equation}
\Pi(Q^{2})|_{V-A}\;=-\frac{32\pi}{9}\;\frac{\alpha_{s}<\bar{q}q>^{2}}{Q^{6}%
}\left\{  1+\frac{\alpha_{s}(\mu^{2})}{4\pi}\left[  \frac{244}{12}%
+\mathrm{ln}\left(  \frac{\mu^{2}}{Q^{2}}\right)  \right]  \right\}
+O(1/Q^{8})\;, \label{2.5}%
\end{equation}
in the $\overline{MS}$ scheme, and assuming vacuum saturation of the
four-quark condensate. Radiative corrections for $d\geq8$ are not known. To
facilitate comparison with current conventions in the literature it will be
convenient to absorb the Wilson coefficients, including radiative corrections,
into the operators, and rewrite Eq.(\ref{2.4}) compactly as

\begin{equation}
\Pi(Q^{2})=\sum_{N\geq3}^{\infty}\frac{1}{Q^{2N}}\;\mathcal{O}_{2N}%
(Q^{2})\;,\label{2.51}%
\end{equation}
where we are dropping the subscript (V-A) for simplicity. We will be concerned
with Finite Energy Sum Rules (FESR), which are nothing but the Cauchy
integral
\begin{equation}
\frac{1}{4\pi^{2}}\int_{0}^{s_{0}}ds\text{\thinspace}P_{N}(s)\left[
v(s)-a(s)\right]  -f_{\pi}^{2}P_{N}(m_{\pi}^{2})=-\frac{1}{2\pi i}%
\oint_{|s|=s_{0}}ds\text{\thinspace}P_{N}(s)\;\Pi^{QCD}\left(  s\right)
\;,\label{2.52}%
\end{equation}
where $P_{N}(s)$ is an arbitrary polynomial, i.e.
\begin{equation}
P_{N}(s)=a_{0}+a_{1}s+a_{2}s^{2}+\ldots+a_{N}s^{N},\label{POLY}%
\end{equation}
$f_{\pi}=92.4\pm0.26\text{ MeV}$ \cite{PDG}, and $v(s)$ ($a(s)$) is the vector
(axial-vector) spectral function measured by ALEPH in tau-decay \cite{ALEPH2},
normalized to the asymptotic value
\begin{equation}
v(s)_{QCD}=a(s)_{QCD}=\frac{1}{2}\left(  1+\frac{\alpha_{s}}{\pi}+...\right)
\;.\label{2.310}%
\end{equation}
The axial-vector spectral function $a(s)$ does not include the pion pole
contribution which is added separately. For most purposes one can work in the
chiral limit $m_{\pi}=0$, i.e. $P_{N}(m_{\pi}^{2})$ in Eq.(\ref{2.52}) may be
replaced by $a_{0}f_{\pi}^{2}$. The standard FESR follow from the
theorem of residues and assuming the Wilson coefficients are just numbers,
\begin{equation}
(-)^{(N+1)}\;\mathcal{O}_{2N}(s_0)=\frac{1}{4\pi^{2}}\int_{0}^{s_{0}}%
ds\,s^{N-1}\;[v(s)-a(s)]-f_{\pi}^{2}\;\delta_{N1}%
\ \ \ \  (N=1,2,3...)\;,\label{2.53}%
\end{equation}
where the index $N$ has been rearranged for convenience. 
Strictly speaking, Eq.(\ref{2.53})
only holds for the constant terms of the Wilson coefficients. Otherwise condensates of lower
or higher dimension get mixed when taking into account
radiative corrections due to the logarithmic terms ($\mathrm{ln}(\mu^{2}/Q^2)$ or
higher). However this mixing of operators of different dimensions occurs
only at order $\alpha_{s}^{2}$ in a given FESR \cite{MIX}. For dimensional
reasons the contribution of the operators of higher dimension in Eq.(\ref{2.53}) vanishes for
large $s_{0}$, while that of operators of lower dimension increases with
$s_{0}$. The latter contribution is particularly disturbing for operators of
high dimension. As the logarithmic terms of the relevant Wilson coefficients
are not known (except the one for $\mathcal{O}_{6}$) we will neglect the
contribution of operators of dimension unequal to $2N$ in Eq.(\ref{2.53}). This
approximation is inherent to every sum rule analysis of the $\tau$-data and
can only be justified \emph{a posteriori} by demonstrating that the right hand
side of Eq.(\ref{2.53}) is (almost) constant. We will, however, examine the
order of magnitude of the mixing to be expected by using Eq.(\ref{2.5}) to
estimate the effect of the radiative corrections of $\mathcal{O}_{6}$.

For $N=1,2$ Eq.(\ref{2.53}) leads to the first two (Finite Energy) Weinberg
sum rules, while for $N=3,4$ the sum rules project out the $d=6$ and $d=8$
vacuum condensates, respectively (notice that in the chiral limit
$\mathcal{O}_{2}=\mathcal{O}_{4}=0$). In order to check the convergence of the
sum rules we consider the first Weinberg sum rule
\begin{equation}
W_{1}(s_{0})\equiv\;\frac{1}{4\pi^{2}}\int_{0}^{s_{0}}ds\,[v(s)-a(s)]=f_{\pi
}^{2} \label{2.7}%
\end{equation}
Strictly speaking $s_{0}\rightarrow\infty$, but precocious scaling would imply
that the sum rule should be saturated at moderate values of $s_{0}$. From
Fig.2, which shows $W_{1}(s_{0})$, one can see that this is clearly not the
case, even at the highest energies accessible in $\tau$-decay.
\begin{figure}
[h]
\begin{center}
\includegraphics[
height=2.1801in,
width=2.8177in
]%
{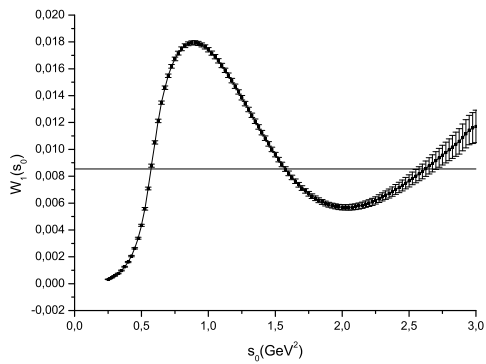}%
\caption{The first Weinberg sum rule, Eq.(\ref{2.7})\thinspace\ as a function
of $s_{0}$ together with $f_{\pi}^{2}=(0.00854\pm0.00005)\; \mbox{GeV}^{2}$ (solid
line).}%
\end{center}
\end{figure}
This lack of precocious scaling in the sum rule can be simply explained by
looking at the measured spectral function in Fig.1. If the spectral function
had reached its approximate asymptotic value (i. e. zero) starting, let us
say, from $s=2\;\mbox{GeV}^{2}$ then the spectral integral of Eq.(\ref{2.7})
would have yielded $f_{\pi}^{2}$ for all $s_{0}\geq2\;\mbox{GeV}^{2}$. This
observation shows us a way out of the dilemma by turning to the more general
sum rules of Eq.(\ref{2.52}). One can choose the polynomial $P_{N}(s)$ in the
sum rule Eq.(\ref{2.52}) in such a way that the problematic contribution of
the integration region near the endpoint of the physical cut is minimized.
This method addresses two problems at the same time, the first being that
experimental errors of the spectral functions grow considerably near the limit
of phase space, and the second, that the asymptotic QCD formula is unreliable
on the contour region near the physical cut. In following this method we will
employ two types of sum rules, pinched FESR and minimizing polynomial sum rules.

\section{Pinched sum rules}

We begin by considering a linear combination of the first two Weinberg sum
rules
\begin{equation}
\bar{W}_{1}(s_{0})\equiv\frac{1}{4\pi^{2}}\int_{0}^{s_{0}}ds\;(1-\frac
{s}{s_{0}})\;[v(s)-a(s)]=f_{\pi}^{2}. \label{3.1}%
\end{equation}
The left hand side of Eq.(\ref{3.1}) as a function of $s_{0}$ is shown in Fig.
3, together with the right hand side.
\begin{figure}
[h]
\begin{center}
\includegraphics[
height=2.2731in,
width=2.9855in
]%
{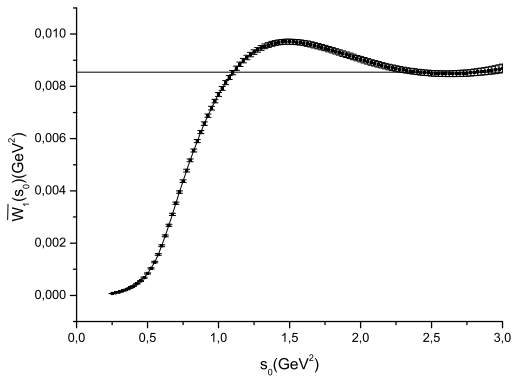}%
\caption{Pinched Weinberg sum rule, Eq.(\ref{3.1}), as a function of $s_{0}$.
The solid line is $f_{\pi}^{2}=(0.00854\pm0.00005)\;\mbox{GeV}^{2}$.}%
\label{Fig.2}%
\end{center}
\end{figure}
It is very reassuring that the sum rule appears to be saturated for
$s_{0}>2.3\;\mbox{GeV}^{2}$. We note that the error band is about a factor
three smaller than that found in a similar analysis \cite{DS2} using the old
ALEPH or OPAL data \cite{ALEPH}-\cite{OPAL}. The influence of the logarithmic 
dependence of $\mathcal{O}_{6}$ in this sum rule is about $4 \times 10^{-6} 
\, {\rm GeV}^2$ in the region of the saturation with $s_0$.

Motivated by this success, we impose the Weinberg sum rules as constraints in
other pinched finite energy sum rules involving different moments. To be
precise, we assume that there are no operators of dimension $d =2$ nor $d =4$, 
which is true in the chiral limit, together with the condition that the FESR
involves factors of $(1-\frac{s}{s_{0}})$ so as to minimize the contribution
near the cut. In this way, we write the Das-Mathur-Okubo \cite{DMO} sum rule in the
form
\begin{equation}
\bar{\Pi}(0)=\frac{1}{4\pi^{2}}\int_{0}^{s_{0}}\frac{ds}{s}\;(1-\frac{s}%
{s_{0}})^{2}\;[v(s)-a(s)]+\frac{2f_{\pi}^{2}}{s_{0}}\;,\label{3.11}%
\end{equation}
where $\bar{\Pi}(0)$ is the finite remainder of the chiral correlator at zero
momentum. It is related to $\bar{L}_{10}$, the counter term of the $O(p^{4})$
Lagrangian of chiral perturbation theory \cite{GL}, which has been calculated independently,

\begin{equation}
\bar{\Pi}(0)=-4 \;\bar{L}_{10}=\left[  \frac{1}{3}f_{\pi}^{2}<r_{\pi}%
^{2}>-F_{A}\right]  \;=0.026\pm0.001\;, \label{3.12}%
\end{equation}

where $<r_{\pi}^{2}>$ is the electromagnetic mean squared radius of the pion,
$<r_{\pi}^{2}>=0.439\pm0.008\;\mbox{fm}^{2}$ \cite{AMEN}, and $F_{A}$ is the
axial-vector coupling measured in radiative pion decay, $F_{A}=0.0058\pm
0.0008$ \cite{PDG}. From Fig. 4 we see that this sum rule is even more
remarkably satisfied. This can be understood by noting that the sum rule
emphasizes less the high-$s$ region where duality violation competes against
stability.
\begin{figure}
[h]
\begin{center}
\includegraphics[
trim=-0.005923in 0.000000in 0.005923in 0.000000in,
height=2.1694in,
width=2.6675in
]%
{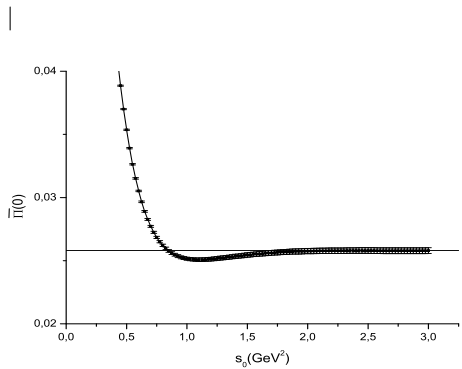}%
\caption{The finite remainder of the chiral correlator at zero momentum,
$\bar{\Pi}(0)$, from Eq.(\ref{3.11}) as a function of $s_{0}$. The solid line is
the central value $\bar{\Pi}(0)=0.02579$.}%
\end{center}
\end{figure}
Numerically, we find from the DMO sum rule
\begin{equation}
\bar{\Pi}(0)=0.02579\pm0.00023\;, \label{3.13}%
\end{equation}
a result showing a remarkable accuracy for a strong interaction parameter. In this
particular sum rule the contribution from the logarithmic term of $\mathcal{O}_{6}$
vanishes.

Next, we turn our attention to the extraction of the condensates with the help
of the pinched sum rules. The philosophy of our calculation is threefold, viz.
(i) to assume that dimension $d =2$ and $d = 4$ operators are absent in the OPE of the
chiral current , (ii) to require that the polynomial projects out only one
operator of the OPE at a time, and (iii) to require that the polynomial and
its first derivative vanish on the integration contour of radius $|s|=s_{0}$.
For the caveats on point (ii) of this approach see the text above. In this way
one obtains for $N\geq 3$ the sum rules (ignoring the energy dependence
in the Wilson coefficients)
\begin{align}
\mathcal{O}_{2N}(s_0) &  =(-1)^{N-1}\Bigg \{\frac{1}{4\pi^{2}}\int_{0}^{s_{0}%
}ds[(N-2)s_{0}^{N-1}-(N-1)s_{0}^{N-2}s+s^{N-1}]\Bigg.\nonumber\\
& \Bigg. \times [v(s)-a(s)] -(N-2)s_{0}^{N-1}f_{\pi}^{2}\Bigg \}\;\;\;\;\;\;\;(N\geq
3)\;.\label{3.2}%
\end{align}
Note that there is always a pinch factor $(s-s_{0})^{2}$ in the polynomial. We
use once again the new ALEPH spectral function and error correlations in this
sum rule. The crucial point of the extraction of the condensates is a careful
inspection of the stability of the result with respect to the variation of all
parameters in the analysis. In our case there is only one parameter, namely
the radius $s_{0}$. This fact contrasts positively with other approaches based
on Laplace sum rules which involve at least two parameters, and in addition do
not project out just one single operator, even for $s_{0}\rightarrow\infty$.
As for stability, if a meaningful value of $\mathcal{O}_{2N}$ is to be
extracted we would expect the r.h.s. of Eq.(\ref{3.2}) to be constant for all
$s_{0}$ larger than some some critical value . We can call this requirement
\textbf{strong stability}. It is best discussed on the basis of the figures
below which show the predictions for various condensates. Figure 5 shows the
result for the dimension $d=6$ condensate.
\begin{figure}
[h]
\begin{center}
\includegraphics[
height=2.1801in,
width=2.694in
]%
{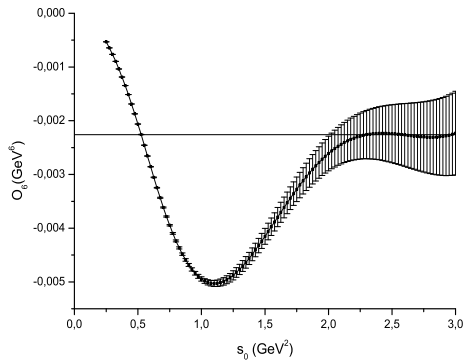}%
\caption{The dimension $d = 6$ condensate, $\mathcal{O}_{6}$, from Eq.(\ref{3.2})
 as a function of $s_{0}$. The solid line is the central value
$\mathcal{O}_{6}=-0.00226 \; \mbox{GeV}^{6}$.}%
\end{center}
\end{figure}
\begin{figure}
[hptb]
\begin{center}
\includegraphics[
height=2.1237in,
width=2.6384in
]%
{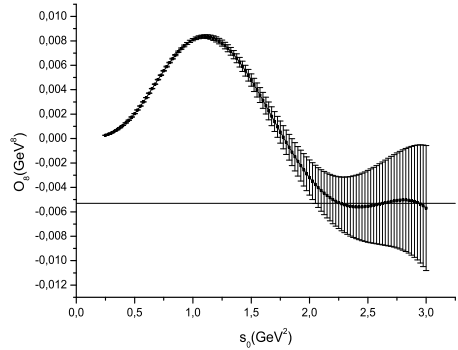}%
\caption{The dimension $d =8$ condensate, $\mathcal{O}_{8}$, from Eq.(\ref{3.4} )
 as a function of $s_{0}$. The solid line is the central value
$\mathcal{O}_{6}=-0.0053\; \mbox{GeV}^{8}$.}%
\end{center}
\end{figure}
There is an obvious stability region: $2.3\leq s_{0}(\mbox{GeV}^{2})\leq3$
from where we find
\begin{equation}
\mathcal{O}_{6}(2.7 \;\mbox{GeV}^{2})=-(0.00226\pm 0.00055)\;\mbox{GeV}^{6}\;.\label{3.3}%
\end{equation}
This value is consistent with the one found from the vacuum saturation
approximation $\mathcal{O}_{6}^{\text{VS}}=-0.0020\;\mbox{GeV}^{6}$ from
Eq.(\ref{2.5}) with $<\bar{q}q>(s_{0})=-0.019\; \mbox{GeV}^{3}$, and $\alpha(s_{0}%
)/\pi=0.1$ , but it is significantly lower than the one found in some earlier
analyses based on the old, incomplete ALEPH data; e.g. $\mathcal{O}%
_{6}=-(0.004\pm0.001)\;\mbox{GeV}^{6}$, obtained in \cite{DS2} using a similar
stability criterion as here. The contribution of the logarithmic term from the 
$\mathcal{O}_{6}$ coefficient in (\ref{3.3}), in the region of $s_0$ considered here, is about
$8 \times 10^{-6}\;\mbox{GeV}^{6}$ and hence negligible within the errors.

To facilitate a comparison with an alternative
type of sum rules discussed in the next section, we give the sum rule for
$\mathcal{O}_{8}$ explicitly
\begin{equation}
\mathcal{O}_{8}(s_{0})=-\frac{1}{4\pi^{2}}\int_{0}^{s_{0}}ds\;[2s_{0}%
^{3}-3s_{0}^{2}s+s^{3}][v(s)-a(s)]+2s_{0}^{3}f_{\pi}^{2}\;.\label{3.4}%
\end{equation}
(Notice that 
$2s_{0}^{3}-3s_{0}^{2}s+s^{3}=\left(  s_{0}-s\right)^{2}\left(s+2s_{0}\right)$).
The result for $\mathcal{O}_{8}$ is shown in Fig.6. In
spite of the larger errors there is still a distinct region of duality in the
interval: $2.3\leq s_{0}(\mbox{GeV}^{2})\leq3\;$, which yields
\begin{equation}
\mathcal{O}_{8}(2.6\; \mbox{GeV}^{2})=-(0.0054\pm0.0033)\;\mbox{GeV}^{8}\label{3.5}%
\end{equation}
Both the sign and the numerical value of this condensate are controversial
(see e.g. Table 1 in \cite{Friot}). Our result agrees within errors with e.g.
that of \cite{Bijnens}-\cite{CGM}, and that of \cite{Rojo} but disagrees in
sign with \cite{Narison}, \cite{DGHS}, \cite{Ioffe}, and with the result of
the minimal hadronic approximation of large $N_{c}$ \cite{Friot}. In
\cite{Narison} the sum rules were evaluated at very low values of $s_{0}$,
mainly because the data at that time was considered to be too inaccurate at
higher $s_{0}$. Fortunately, this state of affairs has now changed with the
new ALEPH analysis. Eq.(\ref{2.5}) can be used to estimate the effect on
$\mathcal{O}_{8}$ due to the mixing of $\mathcal{O}_{6}$ arising from the
logarithmic term. We find a correction of $\ $about$-3 \times10^{-4}\; \mbox{GeV}^{8}$
which is negligible compared to the data error in Eq.(\ref{3.5}). The results
of the sum rules for $\mathcal{O}_{10}$, and $\mathcal{O}_{12}$ as given by
Eq.(\ref{3.2}) are shown in Figs. 7 and 8, respectively.
\begin{figure}
[hptbptb]
\begin{center}
\includegraphics[
height=2.1694in,
width=2.6019in
]%
{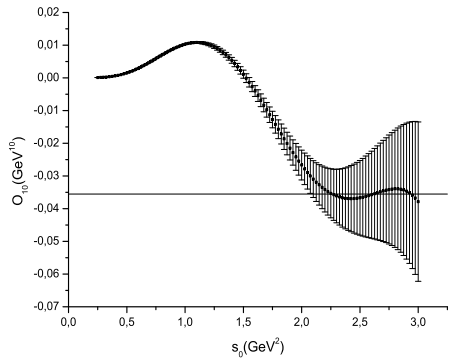}%
\caption{The dimension $d = 10$ condensate, $\mathcal{O}_{10}$, from Eq.(\ref{3.2} )
 as a function of $s_{0}$. The solid line is the central value
$\mathcal{O}_{10}=0.0355 \; \mbox{GeV}^{10}$.}%
\end{center}
\end{figure}
The strong stability obtained so far is now no longer obvious, and we find at
$s_{0}\approx 2.5 \;\mbox{GeV}^{2}$
\begin{align}
\mathcal{O}_{10} (2.5\; \mbox{GeV}^2)&  =(0.036\pm 0.014)\;\mbox{GeV}^{10}\label{3.6}\\
\mathcal{O}_{12}(2.5\; \mbox{GeV}^2) &  =-(0.12\pm 0.05)\;\mbox{GeV}^{12}\;.\label{3.7}%
\end{align}
These results, though, should be taken\emph{\ }\textit{cum}
\textit{grano} \textit{salis}, to wit. Because the OPE is an asymptotic
series, the upper limit of the integration range, $s_{0}$, should increase as
the dimension of the operators increases. The comparison of the numerical
results for $\mathcal{O}_{6}$, $\mathcal{O}_{8}$, $\mathcal{O}_{10}$, and
$\mathcal{O}_{12}$ indicates that at a scale of about $|s|=1\;\mbox{GeV}^{2}$
the OPE starts to diverge at dimension $d=10$. In addition, the problem of the
mixing of operators of different dimensions becomes more severe for higher
dimensional operators. For these reasons we believe that it is rather
meaningless to extract quantitative results for condensates of dimension
higher than $d=8$ from $\tau$-decay spectral functions.
\begin{figure}
[hptbptbptb]
\begin{center}
\includegraphics[
height=2.1694in,
width=2.6492in
]%
{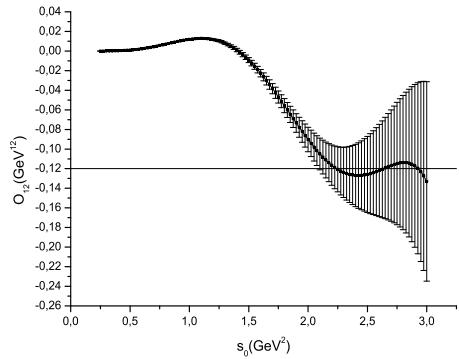}%
\caption{The dimension $d = 12$ condensate, $\mathcal{O}_{12}$, from Eq.(\ref{3.2} )
 as a function of $s_{0}$. The solid line is the central value
$\mathcal{O}_{12}=-0.12 \;\mbox{GeV}^{12}$.}%
\end{center}
\end{figure}

\section{Minimizing Polynomial Sum Rules}

The starting point in this analysis is the general sum rule Eq.(\ref{2.52}).
The polynomial can be chosen in such a way that the problematic contribution
of the integration region near the endpoint of the physical cut is minimized.
With the normalization condition
\begin{equation}
P_{N}\left(  s=0\right)  \,=\,1, \label{4.1}%
\end{equation}
we require that the polynomial $P_{N}(s)$ should minimize the contribution of
the continuum in the range $\left[  c,s_{0}\right]  $ in a least square sense,
i.e.
\begin{equation}
\int_{c}^{s_{0}}s^{k}P_{N}(s)\,\,ds=0\,\,\;\;\;\;\; \,\,(k=0,\ldots N-1)\;,
\label{4.2}%
\end{equation}
The parameter $c$ can be chosen freely in the interval $0<c<s_{0}$. On the
basis of the spectral function of Fig.1, a reasonable choice would be $2
\;\mbox{GeV}^{2}\leq c \leq s_{0}\sim3 \;\mbox{GeV}^{2}$. In a sense, these
polynomials are a generalization of pinched moments, and the $P_{N}(s)$ will
approach $(s-s_{0})^{N}$ when $c \rightarrow s_{0}$. While pinched moments
eliminate the contribution on the physical real axis at a single point, our
polynomials tend to eliminate a whole region (from $c$ to $s_{0}$). The degree
$N$ of the polynomial can be chosen appropriately to project out certain terms
in the OPE, Eq.(\ref{2.51}). The polynomials obtained in this way are closely
related to the Legendre polynomials as follows. Let us introduce the variable
\begin{equation}
x(s) \equiv\frac{2 s - (s_{0} +c)}{(s_{0} - c)} = \frac{2s}{(s_{0}-c)} + x(0)
\;, \label{5.1}%
\end{equation}
and define the polynomials as
\begin{equation}
P_{N}(s) = \frac{L_{N}[x(s)]}{L_{N}[x(0)]} \; , \label{5.2}%
\end{equation}
where $L_{N}(x)$ are the Legendre polynomials
\begin{equation}
L_{N}(x) = \frac{1}{2^{N} N!} \frac{d^{N}}{dx^{N}} (x^{2} -1)^{N} \;.
\label{5.3}%
\end{equation}
We give here only the first few minimizing polynomials
\begin{align}
P_{1}\left(  s\right)   &  = 1 - \frac{2 s}{(s_{0} + c)}\label{4.200}\\
P_{2}\left(  s\right)   &  =\frac{3\left(  2s-\left(  s_{0}+c\right)  \right)
^{2}-\left(  s_{0}-c\right)  ^{2}}{3\left(  s_{0}+c\right)  ^{2}-\left(
s_{0}-c\right)  ^{2}}\label{4.21}\\
P_{3}\left(  s\right)   &  =\frac{5\left(  2s-\left(  s_{0}+c\right)  \right)
^{3}-3\left(  s_{0}-c\right)  ^{2}\left(  2s-\left(  s_{0}+c\right)  \right)
}{-5\left(  s_{0}+c\right)  ^{3}+3\left(  s_{0}-c\right)  ^{2}\left(
s_{0}+c\right)  } \label{4.22}%
\end{align}

If the polynomials are expressed as in Eq. (8), then from Eqs.(\ref{2.52}) and
(\ref{2.53}) there follows the sum rule
\begin{equation}
a_{0}d_{0}+a_{1}d_{1}+...+ a_{N}d_{N}-f_{\pi}^{2} =a_{2}\mathcal{O}_{6}%
-a_{3}\mathcal{O}_{8}+...+\left(  -1\right)  ^{N} a_{N}\mathcal{O}_{2N+2} \; ,
\label{4.3}%
\end{equation}
where the constants
\begin{equation}
d_{N}= \frac{1}{4\pi^{2}} \int_{0}^{s_{0}}ds\,s^{N}\,\left(  v(s)-a(s)\right)
\label{4.4}%
\end{equation}
are to be determined from the ALEPH data.

We begin with the $\mathcal{O}_{6}$ condensate and obtain from Eq.(\ref{4.3})
the sum rule
\begin{equation}
d_{0}+a_{1}d_{1}+a_{2}d_{2}-f_{\pi}^{2}=a_{2}\mathcal{O}_{6}\;.\label{4.5a}%
\end{equation}
After substituting $P_{2}$ from Eq.(\ref{4.21}) this sum rule becomes
\begin{equation}
\mathcal{O}_{6}(s_0)=\frac{1}{6}(s_{0}^{2}+4s_{0}c+c^{2})\left(  d_{0}-f_{\pi}%
^{2}\right)  -\left(  s_{0}+c\right)  d_{1}+d_{2}\;.\label{4.7}%
\end{equation}
In the sequel we choose the initial value $s_{0}=3\;\mbox{GeV}^{2}$, but will
subsequently change it in the range $2.5\leq s_{0}(\mbox{GeV}^{2})\leq3$ in
order to verify the criterion of \textbf{strong stability}. The condensate
$\mathcal{O}_{6}$ from Eq.(\ref{4.7}) is plotted in Fig. 9 as a function of
$c$.
\begin{figure}
[h]
\begin{center}
\includegraphics[
height=2.1428in,
width=2.8634in
]%
{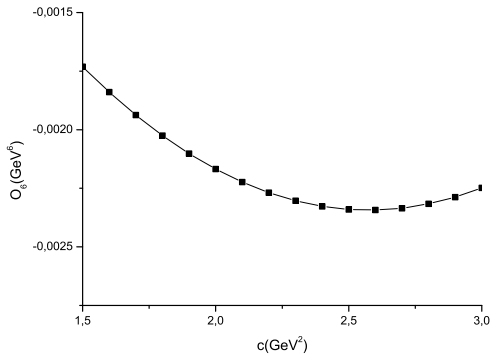}%
\caption{The condensate $\mathcal{O}_{6}$ as a function of the parameter $c$.}%
\label{O6(c)}%
\end{center}
\end{figure}
One can appreciate a stable point for $c=2.5 \;\mbox{GeV}^{2}$. Fixing $c$ at
this point of minimal sensitivity has been discussed previously in other FESR
applications, e.g. in \cite{BPS}. For $c=2.5\;\mbox{GeV}^{2}$, and
$s_{0}=3\;\mbox{GeV}^{2}$ we obtain from Eq.(\ref{4.7}) the result:
\begin{equation}
\mathcal{O}_{6}(3\; \mbox{GeV}^{6})=-(0.0023\pm0.0013)\;\mbox{GeV}^{6}\;,\label{4.14}%
\end{equation}
which compares well within errors with the previous result from the pinched
sum rule, Eq.(\ref{3.3}). We have tested positively the stability around this
point by choosing different values of $s_{0}$ in the range $2.5\leq
s_{0}(\mbox{GeV}^{2})\leq3$. For instance, using $s_{0}=2.5\;\mbox{GeV}^{2}$
we obtain $O_{6}(2.5 \;\mbox{GeV}^{6})=-(0.00224\pm 0.00046)\;\mbox{GeV}^{6}$. Other
values of $s_{0}$ in the above range lead to similar results, well in
agreement within errors with Eq.(\ref{4.14}), thus satisfying the criterion of
\textbf{strong stability}.

Next, we consider the $\mathcal{O}_{8}$ condensate, and use Eq.(\ref{4.3}) to
obtain the sum rule
\begin{equation}
d_{0}+a_{1}d_{1}+a_{2}d_{2}+a_{3}d_{3}-f_{\pi}^{2}=a_{2}\mathcal{O}_{6}%
-a_{3}\mathcal{O}_{8}\;.\label{4.15}%
\end{equation}
The presence of $\mathcal{O}_{6}$ in the sum rule for $\mathcal{O}_{8}$ can be
dealt with in two ways. One could insert the numerical value of $\mathcal{O}%
_{6}$, e.g. Eq.(\ref{4.14}), as obtained from its own sum rule, or rather
substitute the analytic expression of the sum rule itself. The latter
procedure yields the best possible results in terms of stability and accuracy,
and leads to the sum rule
\begin{equation}
\mathcal{O}_{8}(s_0)=-\frac{1}{5}\;[s_{0}(s_{0}+2c)^{2}+c^{3}]\;(d_{0}-f_{\pi}%
^{2})+\frac{3}{10}\;[3(s_{0}^{2}+c^{2})+4s_{0}c]\;d_{1}-d_{3}\;. \label{4.16}
\end{equation}
Notice the welcome absence of the term involving the second moment, i.e.
$d_{2}$; it cancels out when substituting in Eq.(\ref{4.15}) the sum rule for
$\mathcal{O}_{6}$, Eq.(\ref{4.7}). Choosing again the initial value
$s_{0}=3\;\mbox{GeV}^{2}$, one obtains for $\mathcal{O}_{8}$ the results shown
in Fig. 10. One can see again a stability region near $c=2.5\;\mbox{GeV}^{2}$,
leading to the result
\begin{equation}
\mathcal{O}_{8}(3 \;\mbox{GeV}^{2})=-(0.0048\pm 0.0039)\;\mbox{GeV}^{8}%
\end{equation}
It is worth mentioning that the polynomial coefficients entering the sum rule
Eq.(\ref{4.16}) differ significantly from the ones in the corresponding pinched
sum rule Eq.(\ref{3.4}). It is therefore reassuring that both results for
$\mathcal{O}_{8}$ are compatible. To check for \textbf{strong stability} we
have, once again, varied $s_{0}$ in the range $2.5\leq s_{0}(\mbox{GeV}^{2}%
)\leq3$. For $s_{0}=2.5\;\mbox{GeV}^{2}$ we find $O_{8}(2.5 \; \mbox{GeV}^{8}
)=-(0.0056\pm0.0024)\;\mbox{GeV}^{8}$, and similar results for other values of
$s_{0}$ in the above range. Thus, the criterion of \textbf{strong stability}
is again satisfied, albeit within large errors. Proceeding beyond dimension
$d=8$ is marred by the same problems mentioned at the end of Section 3; the
minimizing polynomial FESR do not avoid the divergence of the OPE, nor the
increasing importance of operator mixing.
\begin{figure}
[h]
\begin{center}
\includegraphics[
height=2.1619in,
width=2.8634in
]%
{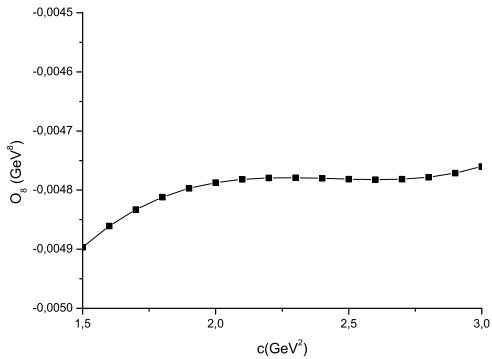}%
\caption{$\mathcal{O}_{8}$ as a function of the parameter $c$.}%
\end{center}
\end{figure}

\section{Conclusion}

The final ALEPH data for the chiral spectral function $v(s)-a(s)$ shows
clearly that this spectral function has not yet reached its asymptotic form
dictated by perturbative QCD, i.e. it does not vanish, even at the highest
energies attainable in $\tau$-decay. If the asymptotic regime had been reached
precociously, let us say at $Q^{2}\simeq2\;\mbox{GeV}^{2}$, then it would have
been straightforward to calculate the non-perturbative condensates with the
help of the Cauchy Integral. Since this is not the case, some method to
improve convergence must be applied. We have shown that in the framework of
FESR this can be done by suitably reducing the impact of the high energy
region in the dispersive integral, either by using pinched sum rules or by
using minimizing polynomial sum rules. We first used the data in a pinched
linear combination of the first two Weinberg sum rules which follow from the
fact that there are no dimension $d =2$ and $d =4 $ operators contributing to the chiral
correlator to demonstrate the precocious saturation of the sum rule and the
remarkable effectiveness of the method. Motivated by this success, we
determined a number of QCD condensates by making maximal use of the fact that
there are no dimension $d =2$ and $d =4$ operators and requiring \textbf{strong
stability} for both methods, i.e. we varied the radius $s_{0}$ in the Cauchy
integral beginning at the end of $\tau$-decay phase space and required that
the condensates calculated from the data should be reasonably constant for all
$s_{0}$ in some finite region including the end of phase space. We do not
assume (as is done in most FESR calculations) that the dispersive integral
vanishes from $s_{0}\rightarrow\infty$. By showing that there is "strong
stability" i.e. precocious saturation of the FESR we prove that this region
contributes only negligibly. It would  indeed be surprising if the observed
stability would disappear for $s_{0}$ larger than the end of phase space. We
do, however, have to make the assumption, inherent in all sum rule analyses of
$\tau$-decay, that unknown $O(\alpha_{s}^{2})$ effects of mixing of operators
of different dimensions are negligible for the relevant duality range
$2.5 \;\mbox{GeV}^{2}\lesssim s_{0}\lesssim 3\; \mbox{GeV}^{2}$. We have checked 
explicitely that in all the cases considered in this work this is the situation
when one takes into account the logarithmic term of the 
dimension six Wilson coefficient. The results for $\mathcal{O}_{6}$
and $\mathcal{O}_{8}$ satisfy this strong stability criterion as is best seen
from the figures. Extraction of higher condensates of dimension $d\geq10$
leave room for interpretation, but the conclusion that they grow rapidly with
dimensionality is rather obvious. Together with the increasing importance of
operator mixing, it makes the extraction of these condensates a difficult
exercise. Our result that $\mathcal{O}_{6}$ and $\mathcal{O}_{8}$ have the
same sign is in conflict with some of the earlier determinations based on the
incomplete ALEPH data but agrees with others (see e.g. \cite{Friot} for a
comparative summary).

ACKNOWLEDGMENTS: We wish to thank Hubert Spiesberger and Alexei Pivovarov for discussions.


\begin{thebibliography}{99}                              
\bibitem {SVZ}M.A. Shifman, A.I. Vainshtein, and V.I. Zakharov, Nucl. Phys.
\textbf{B 147} (1979) 385.

\bibitem {Friot}S. Friot, D. Greynat and E. de Rafael, J. High Energy Phys.
\textbf{0410} (2004) 043

\bibitem {ALEPH2}ALEPH Collaboration, R. Barate \textit{et al.},
hep-ex/0506072 (2005).

\bibitem {WSR}S. Weinberg, Phys. Rev. Lett. \textbf{18} (1967) 507;

\bibitem {DMO}T. Das, V.S. Mathur, and S. Okubo, Phys. Rev. Lett. \textbf{19}
(1967) 859.

\bibitem {Nasrallah}N.F. Nasrallah, N.A. Papadopoulos and K. Schilcher, Phys.
Lett. \textbf{B 126} (1983) 1983; C.A. Dominguez, K. Schilcher, Phys. Lett.
\textbf{B 448} (1999) 93.

\bibitem {BPS}J. Bordes, J. Pe\~{n}arrocha, K. Schilcher, J.High Energy Phys.
\textbf{0412} (2004) 064.

\bibitem {CH}L.\ V.\ Larin, V.\ P.\ Spiridonov, K.\ G.\ Chetyrkin, Sov. J.
Nucl. Phys. \textbf{44} (1986) 892.

\bibitem {Bijnens}J. Bijnens, E. Gamiz and J. Prades, J.High Energy Phys.
\textbf{0110} (2001) 009

\bibitem {CGM}V. Cirigliano, E. Golowich, K. Maltman, Phys. Rev. \textbf{D 68}
(2003) 054013.

\bibitem {Cirigliano}V. Cirigliano, J. F. Donoghue, E. Golowich, K. Maltman,
Phys.Lett. B522 (2001) 245

\bibitem {PDG}Particle Data Group, S. Eidelman et al., Phys. Lett. \textbf{B
592} (2004) 1.

\bibitem {MIX}G. Launer Z. Physik \textbf{C 32} (1986) 557.

\bibitem {DS2}C.A. Dominguez, K. Schilcher, Phys. Lett..\textbf{B 581} (2004) 193.

\bibitem {ALEPH}ALEPH Collaboration, R. Barate \textit{et al.}, Eur. Phys. J.
\textbf{C 4} (1998) 409.

\bibitem {OPAL}OPAL Collaboration, K. Ackerstaff \textit{et al.}, Eur. Phys.
J. \textbf{C 7} (1999) 571; G. Abbiendi \textit{et al.}, \textit{ibid.}
\textbf{C 13} (2000) 197.

\bibitem {GL}J. Gasser and H. Leutwyler, Nucl. Phys. \textbf{B 250} (1985)
465; G. Ecker, J. Gasser, A. Pich, and E. de Rafael, Nucl. Phys. \textbf{B
321} (1989) 311.

\bibitem {AMEN}S.R. Amendolia et al., Nucl.Phys. \textbf{B 277} (1986) 168.

\bibitem {Rojo}J. Rojo, J.I. Latorre, J.High Energy Phys. \textbf{0401} (2004) 055.

\bibitem {Narison}S. Narison, hep-ph/0412152 (2005). This paper contains
extensive reference to earlier work.

\bibitem {DGHS}M.\ Davier, L.\ Girlanda, A.\ H\"{o}cker, J.\ Stern,
Phys.\ Rev.\ \textbf{D 58} (1998) 096014.

\bibitem {Ioffe}B.\ L.\ Ioffe, K.\ N.\ Zyablyuk, Nucl.\ Phys.\ \textbf{A 687} (2001) 437. The sign of $\mathcal{O}_{8}$ has been changed in later publications, see e.g. B.L. Ioffe, Progr. Part. Nucl. Phys. \textbf{56} (2006) 232.

\end{thebibliography}
\end{document}